\documentclass{llncs}
\usepackage{graphicx}

\usepackage{amsmath}
\usepackage{amssymb}
\usepackage{algorithm}
\usepackage{algorithmic}
\begin{document}

\title{{\sc PolyDNN}\\ Polynomial Representation of NN for Communication-less SMPC Inference} 
\author{Philip Derbeko and Shlomi Dolev}
\institute{Department of Computer Science, Ben-Gurion University of the Negev,\\
Beersheba, Israel}
\date{October 2020}
\maketitle

\begin{abstract}
    The structure and weights of Deep Neural Networks (DNN) typically encode and contain very valuable information about the dataset that was used to train the network. 
    One way to protect this information when DNN is published is to perform an interference of the network using secure multi-party computations (MPC). 
    In this paper, we suggest a translation of deep neural networks to polynomials, which are easier to calculate efficiently with MPC techniques. 
    We show a way to translate complete networks into a single polynomial and how to calculate the polynomial with an efficient and information-secure MPC algorithm.    
    The calculation is done without intermediate communication between the participating parties, which is beneficial in several cases, as explained in the paper.
\end{abstract}

\section{Introduction}
Deep Neural Networks (DNN) are the state-of-the-art form of Machine Learning techniques these days. 
They are used for speech recognition, image recognition, computer vision, natural language processing, machine translation, and many other tasks. 
Similar to other Machine Learning (ML) methods, DNN is based on finding patterns in the data and, as such, the method embeds information about the data into a concise and generalized model.
Subsequently, the sharing of the DNN model also reveals private and valuable information about the data. 

In this paper, we consider a situation where the data owner collects the data, trains the model, and shares the model to be used by clients.
The model is very valuable for the data owner, as the training process is resource-intensive and frequently performed over private and valuable data. 
Therefore, the goal of the data owner is to retain control of the model as much as possible after it was shared. 
The data owner will likely be willing to delegate the query service to (Machine Learning Data model Store MLDSore \cite{DBLP:conf/cscml/DerbekoDG19}) clouds, in a way that the cloud providers do for computing platforms.   
In such a scenario, the cloud providers should not be able to simply copy the model and reuse it.
In addition, the data owner should have the ability to limit the number of queries executed on the model, such that a single, or a small team of colluding cloud providers (or servers) cannot execute an unlimited number of queries on the model. In fact, we show that a (D)NN can be represented by a (nested) polynomial, therefore enough queries (points on the polynomial) can reveal the neural network, and the ownership of the information (succinct model) is at risk. In practice, as data is constantly updated new data emerges and old data becomes obsolete, thus, frequent updates of the neural network are frequent enough to define a new polynomial for which the past queries are not relevant.   

In our previous work~\cite{DBLP:conf/bigdataconf/DerbekoDG19} we have shown an algorithm to distributively share DNN-based models while retaining control and ownership over the shared model. 
The activation functions of neural network units were approximated with polynomials and we have shown an efficient, additive secret sharing based, MPC protocol for information-secure calculations of polynomials. 

In this paper, we first suggest approximating a trained neural network with a single (possibly nested) polynomial. 
We present a nested polynomial approach to speed up the calculation of the polynomial on a single node. 
The essence of the idea is to nest the polynomial approximation of each layer within the approximation of the next layer, such that a single polynomial (or arithmetic circuit) will approximate not only a single network unit, but a few layers or even the entire network.
We discuss an efficient, (perfect information theoretically secure) secret-sharing MPC calculation of the polynomial calculation of DNN.
Lastly, we compare the MPC calculation of the neural network itself with a calculation of polynomial representation. 

Our main contribution in this research is an optimization of (communication-less) MPC calculations of a shared DNN by approximating neighboring layers by a single polynomial, and in some cases, the entire network.
An additional contribution is a nesting of a multi-layer polynomial to reduce the redundant calculations of the intermediate layers.

The rest of the paper is structured as follows;
Previous relevant research is covered in Section~\ref{sec:previous work}. 
Section~\ref{sec:polynomial neural network} discusses the most common activation functions and how they are approximated with polynomials.
Section~\ref{sec:multiple layers approximation} discusses a way to approximate DNN with a single polynomial on a single computing node. The premise of the section is to establish a basis for a secure, communication-less multi-party computation, which is presented in Section~\ref{sec:efficient MPC}. Section \ref{sec:communicationless mpc for dnn} summarizes the techniques for blindly computing a 
polynomial (some of its coefficients being secret shares of zero) to obtain blind execution of DNN. 
Section \ref{sec:FHA} describes the way our polynomial neural network representation facilitate an efficient execution of the inference by an untrusted third party, without revealing the (machine learning big) data the queries, and the results. 
Empirical experiments are described in Section~\ref{sec:experiments} and, lastly, the paper is concluded in Section~\ref{sec:conclusions}.

\section{Previous Work}
\label{sec:previous work}
The main issue in calculating the neural network activation with secure multi-party computations algorithms is the translation of activation functions from floating-point arithmetic to fixed-point.
The regular activation functions of neural network units use floating-point arithmetic for the accuracy of the  calculations, as the penalty of such calculations on modern CPUs is small. 
However, distributed MPC protocols are built for fixed-point arithmetic, and many times even for limited range values. 

Approximation of neural network units' activation function with fixed-point arithmetic was considered before in~\cite{GiladBachrach2016CryptoNetsAN,DBLP:journals/corr/abs-1711-05189}, where polynomial functions were suggested for approximation. In both works, the network was not distributed but rather was translated into fixed-point calculations to run over encrypted data. However, the methods of approximation are similar to the MPC case.

CryptoDL~\cite{DBLP:journals/corr/abs-1711-05189} showed an implementation of Convolutional Neural Networks (CNN) over encrypted data using homomorphic encryption (HE). As fully homomorphic encryption is limited to addition and multiplication operations, the paper has shown approximation of CNN activation functions by low-degree polynomials due to the high-performance overhead of higher degree polynomials.

The calculation of neural networks with secure multi-party computations was considered in~\cite{SecureML}. 
Sigmoid and softmax activation functions were replaced with functions that can be calculated with MPC. 
Their experiments showed that the polynomial approximation of the sigmoid function requires at least a 10-degree polynomial, which causes a considerable performance slow-down with garbled circuit protocol. 
Thus, they propose to replace sigmoid and softmax activation functions with a combination of ReLu activation functions, multiplication, and addition.
Our experiments show that the accuracy of polynomial function approximation increases with the degree, $d$, of the polynomial and flattens at about $d=30.$
As a side-note, a polynomial degree is a less important issue in our computation scheme for reasons explained later in the paper. 
Their work had a limitation for two participating parties and the algorithm was shown to be limiting in terms of performance and the practical size of the network. 

CrypTFlow~\cite{kumar2019cryptflow} is a system that converts TensorFlow (TF) code automatically into secure multi-party computation protocol. 
The system has three parts: a compiler, from TF code into two and three-party secure computations, an optimized three-party computation protocol for secure interference, and a hardware-based solution for computation integrity. 
The most salient characteristic of CrypTFlow is the ability to automatically translate the code into MPC protocol, where the specific protocol can be easily changed and added. 
The optimized three-party computational protocol is specifically targeted for NN computation and speeds up the computation. This approach is similar to the holistic approach of~\cite{Agrawal2019QUOTIENTTS}.

SecureNN~\cite{Wagh2018SecureNNEA} proposed arguably the first practical three-party secure computations, both for training and for activation of DNN and CNN. 
The impressive performance improvement over then, state-of-the-art, results is achieved by replacing garbled circuits and oblivious transfer protocols with secret sharing protocols. The replacement also allowed information security instead of computational security. 
The paper provides a hierarchy of protocols allowing calculation of activation functions of neural networks. 
The downside is that the protocols are specialized for three-party computations and their adaptation for more computational parties is not trivial, unlike our scheme that supports any number of participating servers.
Also, despite being efficient, the paper claimed more than a four times speedup over SecureML's~\cite{SecureML} work, the protocols require ten communication rounds for ReLu calculation of a single unit not counting share distribution rounds. 


A different approach at speeding up performance was made by~\cite{Demmler2015ABYA}, which concentrated on two-party protocols.
The work showed a mixed protocol framework based on Arithmetic sharing, Boolean sharing, and Yao's garbled circuit (ABY). Each protocol was used for its specific ability, and the protocols are mixed to provide a complete framework for neural networks activation functions.

MPC computations of DNN were considered in additional works,  for example, in~\cite{Rouhani2018DeepsecureSP,Agrawal2019QUOTIENTTS}, which have a lesser connection to our contribution.

We begin by showing how DNN can be approximated with polynomial functions representing a single, or multiple layers.
As an approximation of multiple layers increases the degree of the polynomial function, we show possible techniques to use (intermediate) communication to keep the degree low in Section~\ref{sec:multiple layers approximation}. 

\section{Neural Network as Polynomial Functions in a Single Node Case}
\label{sec:polynomial neural network}
We show how to approximate functions that are a typical part of DNNs, by polynomials. 
We focus on the most commonly used functions in neural networks. \\

\begin{figure}[!ht]
    \centering
	 \includegraphics[width=0.3\textwidth]{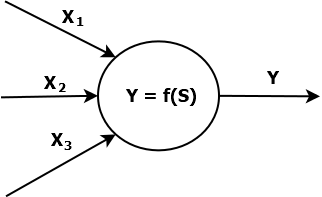}
	 \caption{
		 A schematic representation of a single neuron unit. 
		 The neuron receives inputs $X_1, \ldots, X_n$, calculates a weighted sum of the inputs $S=\sum_{i=1}^n w_i X_i - b.$
		 Finally, the output of the unit is the result of the activation function $f()$ on $S$.
	}
	 \label{fig:neuron}
\end{figure}

\noindent{\bf Weighted Sum of the Unit Input.}
Figure~\ref{fig:neuron} depicts a schema of a single neuron.
Given neuron inputs $X_1, \ldots, X_n$, the weighted sum is a multiplication of inputs with the corresponding weights $S=\sum_{i=1}^n w_i X_i - b,$ where $b$ is a bias of the neuron. 

The sum is very easily approximated with a polynomial, as it is a vector multiplication of weight with the $n$-dimensional input or in other words, a polynomial of degree 1.\\

\noindent{\bf Common Activation Functions.} Most of the research approximating DNN activation functions focused on these few common functions:
\begin{itemize}
    \item ReLu $\left( ReLu(x) = max(0, x) \right)$,
    \item Leaky ReLu (similar to ReLu but $LReLu(x) = 0.01 x$ if $x \leq 0$)
    \item Sigmoid $\left( \sigma(x) = \frac{1}{1+e^{-x}}\right)$
    \item TANh $\left( tanh(x) = \frac{e^{2x}-1}{e^{2x}+1}\right)$
    \item SoftMax (used for multi-class prediction $\sigma(x_i) = \frac{e^{x_i}}{\sum_{i=1}^{k} e^{x_i}}$)
\end{itemize}
All those functions can be approximated with a polynomial using various different methods, for example~\cite{Xie2014CryptoNetsNN,DBLP:journals/corr/abs-1711-05189,SecureML,Agrawal2019QUOTIENTTS,Rouhani2018DeepsecureSP}. 
Our optimization method is agnostic to a specific approximation method. 

Another point to consider is that opposite to most of the research approach in~\cite{DBLP:journals/corr/abs-1711-05189}, which minimized the degree of the approximating polynomial, 
our communication-less approach allows us to use a higher degree polynomials. 
In our previous research~\cite{DBLP:conf/bigdataconf/DerbekoDG19}, that considered a single polynomial for a single neuron unit, rather than a single (nested) polynomial for many or even all units as we do here, we have shown that 30-degree Chebyshev polynomials achieve good results. Increasing the polynomial degree higher results in diminishing returns.\\ 

\noindent{\bf Convolution Layer.}
The convolution layer is used in Convolutional Neural Networks (CNN), mainly for image recognition and classification. Usually, this layer performs dot product of a (commonly) $ n \times n$ square of data points (pixels). The idea is to calculate the local features. The layer performs multiplication and addition, which are directly translated into a polynomial.\\

\noindent{\bf Max and Mean Pooling.}
Max and Mean pooling compute the corresponding functions of a set of units. 
Those functions are frequently used in CNN following the convolution layers. 
Previous works~\cite{Xie2014CryptoNetsNN} suggested replacing max-pooling with a scaled mean-pooling, which is trivially represented by a polynomial. 
However, this requires the replacement to be done during the training stage. 

For networks that did not replace max pooling with mean and as an alternative, max function can be approximated by:
\begin{equation}
    max(x_1, \ldots, x_n) = lim_{d \rightarrow \infty} \left( \sum_i x^d_i \right )^{\frac{1}{d}}
\label{eq:general approximation}
\end{equation}
When $d=1$ the approximation is reduced to a scaled mean pooling function, i.e., without division by the number of elements. 

\begin{figure}[!ht]
    \centering
	 \includegraphics[width=0.7\textwidth]{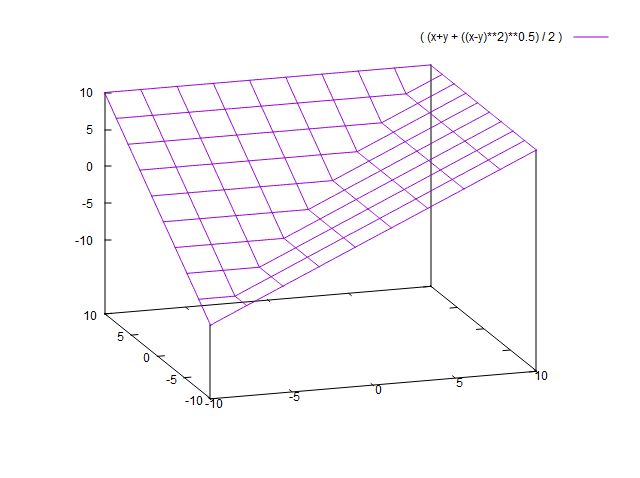}
	 \caption{
		 Equation~\ref{eq:quadratix approximation} of max function.
	}
	 \label{fig:max approximation}
\end{figure}

A simple and practical variation of the equation~\ref{eq:general approximation} is:
\begin{equation}
    m'(x, y) = \frac{x + y}{2} + ((x - y)^2)^{1/2}.
    \label{eq:quadratix approximation}
\end{equation}
Notice that the function provides an approximation near any values of $x$ and $y$, which is an advantage over Taylor or Chebyshev approximations, that are developed according to a specific point. 
Despite its simplicity, equation~\ref{eq:quadratix approximation} provides a relatively good approximation, see Figure~\ref{fig:max approximation}.

Notice that using a two-variable function for the max pooling layer of $k$ inputs requires chaining of the max functions: 
\[ 
max(x_1, x_2, \ldots x_k) = max(x_1, max(x_2, \ldots ,max(x_{k-1}, x_k))).
\]

Alternatively, the optimization sequence is interrupted at the max-pooling layer, which will require an MPC protocol for the max function calculation, for example~\cite{Wagh2018SecureNNEA}.\\

\noindent{\bf Long Short-Term Memory (LSTM).}
LSTM is a subset of Recurrent Neural Network (RNN) architecture, whose goal is to learn sequences of data. 
LSTM networks are used for speech recognition, video processing, time sequences, etc. 

There are many different variations of LSTM units with a usual structure, including several gates or functions, which enable the unit to remember values over several cell activations. 
A common activation function of LSTM units is the logistic sigmoid function, which we already considered in Section~\ref{sec:polynomial neural network}. 

\section{Multiple Layers Approximation}
\label{sec:multiple layers approximation}
We have discussed the approximation of DNN functions by polynomials. 
The approximation exists for all the common functions. 
This makes it possible to combine multiple layers into a single polynomial function according to the connectivity of the layers. 

\begin{figure}[!t]
	 \centering
	 \includegraphics[width=0.7\textwidth]{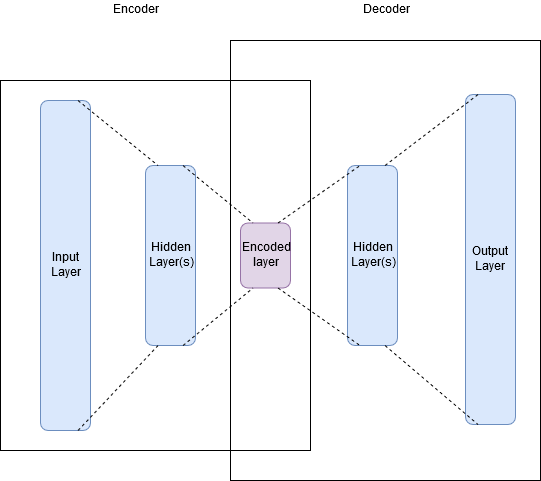}
	 \caption{
		 A high-level architecture of the auto-encoder neural network. 
		 Encoder transforms data from the original dimension to a much smaller encoding, 
		 while the decoder performs the opposite operation of restoring the original data from the encoded representation.
	}
	 \label{fig:autoencoder}
\end{figure}

One example of a network that can be approximated by a single polynomial function is auto-decoder, see Figure~\ref{fig:autoencoder} for a simple example, where hidden layers are dense layers with (commonly) ReLu or sigmoid activation.

\begin{figure}[!ht]
    \centering
	 \includegraphics[width=0.5\columnwidth]{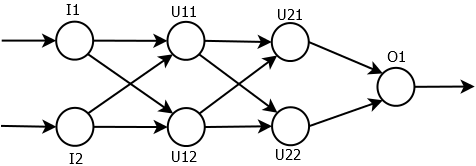}
	 \caption{
		 A small example network with an input layer on the left, two dense hidden layers U1 and U2, and an output layer on the right consisting of a single unit. 
		 Each layer utilizes ReLu or sigmoid activation functions, or any other function that can be approximated by a polynomial.   
	}
	 \label{fig:example poly network}
\end{figure}

The idea is to create a polynomial for the ``flow'' of the data in the network instead of approximating every single neural unit with a polynomial. 
As an example, consider the network in Figure~\ref{fig:example poly network}. 

The network consists of an input layer ($I$) on the left, two dense hidden layers ($U_1$ and $U_2$), and one output layer $O$, which is implemented by the softmax function.
The units are marked as $u_{li}$ where $l$ is the hidden layer number and $i$ is the number of the unit in the layer.
We assume that the activation functions of the hidden layers are ReLu (or any other function that can be approximated by a polynomial function). 

Consider a unit $u_{11}$. It calculates the function which is approximated by the polynomial. Assume that ReLu activation functions are approximated using a polynomial of $d$-degree.
\begin{equation}
ReLu( \sum_i w_i I_i ) \approx P_{11} = Pol_{11}(\sum_i w_i I_i).
\end{equation}
Unit $u_{21}$ receives $P_{11}$ and $P_{12}$ as inputs and calculates the ``nested'' polynomial function:
\begin{equation}
    P_{21} = Pol_{21}(\sum_i w_i P_{1i}).
\end{equation}

In general, assuming dense layers, the nested polynomials are defined as:
\begin{equation}
    \label{eq:general nested poly}
    P_{lj} = Pol_{lj}(\sum_i w_i P_{(l-1)i}).
\end{equation}

In this simple case, the result of networks evaluation can be calculated by evaluating two polynomials of $d^2$-degree: $P_{21}$ and $P_{22},$ and calculating the output layer function of their output. 
Overall, by approximating softmax by $Pol^{sm}$ we get the following polynomial for the entire network:
\begin{equation}
    \begin{tabular}{rl}
    $DNN(x)$ & = $Pol^{sm}$ $\left( w^o_{1} P_{21} + w^o_{2} P_22  \right)$ \\
        = & $Pol^{sm}$ $\left( w^o_{1} Pol_{21} (w^{21}_1 P_{11} + w^{21}_2 P_{12})
        + w^o_{2} Pol_{22} (w^{22}_1 P_{11} + w^{22}_2 P_{12})  \right)$ \\
        = & $Pol^{sm} \left( w^o_{1} Pol_{21} (w^{21}_1 Pol_{11}( w^{11}_1 I_1 + w^{11}_2 I_2) + w^{21}_2 Pol_{12}( w^{12}_1 I_1 + w^{12}_2 I_2)) \right.$ \\
        & + $\left. w^o_{2} Pol_{22} (w^{22}_1 Pol_{11}( w^{11}_1 I_1 + w^{11}_2 I_2) + w^{22}_2 Pol_{12}( w^{12}_1 I_1 + w^{12}_2 I_2))  \right)$
    \end{tabular}
    \label{eq:network polynomial}
\end{equation}

Notice that $P_{11}$ and $P_{12}$ were calculated twice as they are used as inputs for both $U_{21}$ and $U_{22}$ units.\\

\noindent{\bf Non-dense layers} are approximated in a similar way but with only the corresponding inputs from the previous layer. 
An example of such architectures is CNN, commonly used for image recognition.
CNN layers have a {\em topographic structure}, where neurons are associated with a fixed two-dimensional position that corresponds to a location in the input image. 
Thus, each neuron in the convolutional layer is connected (receives its inputs) from a subset of neurons from the previous layer, 
that belong to the corresponding rectangular patch.

For the polynomial approximation of such networks, the polynomial approximating the unit depends only on the relevant units from the previous layer.
In case when the interconnections of the networks, part of the architecture, are part of a secret as well, then a technique from Section~\ref{sec:hiding architecture} can be used. 
In other words, the network is considered to be dense, but the weights of the ``pseudo''-connections are set to zero. 
Thus, achieving the same effect as not connecting the units at all.

\subsection{Nesting to Decrease Polynomial Degree}
\label{sec:nested polynomials}
Neural units' calculation is the most common operation in the DNN feed-forward operation. 
The approximation of the operation with polynomial significantly increases the complexity of the activation function. For instance, ReLu function is in its essence a simple {\em if} condition, and is approximated with a $30-$degree polynomial.

The impact is less severe than expected, as the approximation is applied to the networks after their training phase, which is extremely computationally intensive. 
However, it is still significant if many inference calculations are performed. 
In a multi-layer polynomial approximation, the degree of polynomials increases leading to an increase of performance overhead as well. 

To limit polynomial degree at layer $l$, the polynomials from $l-1$ layer are not explicitly expanded into a single function (see Equation~\ref{eq:general nested poly}), but rather kept as a sum of lower-degree polynomials. 
In this way, each polynomial $P_{ij}$ is calculated only once in the process of the calculation of each multi-layer polynomial. 
This will limit the degree of the polynomial and eliminate redundant calculations.

\subsection{Hiding Network Architecture}
\label{sec:hiding architecture}
In case the network architecture in itself is valuable, it might be desired to conceal the correct architecture from cloud providers. 
The na\"ive multi-layer polynomial approximation hides the network architecture somewhat, even though, the degree of the polynomial might be a telling factor. 

\begin{figure}[!ht]
    \centering
	 \includegraphics[width=0.5\columnwidth]{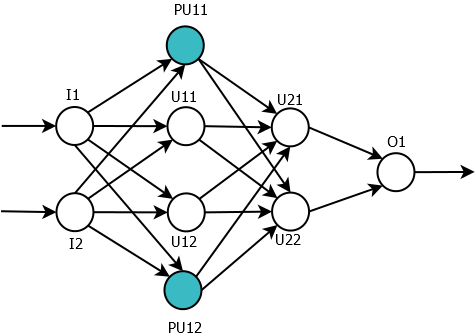}
	 \caption{
		 An example of a network with pseudo-units.
		 The example extends Figure~\ref{fig:example poly network} with two pseudo-units: $PU11$ and $PU12$. 
		 The units are connected just like units of the dense layer, with input from all units of the previous layer and output connected to all units of the next layer.   
	}
	 \label{fig:hidden network example}
\end{figure}

A way to conceal the exact network architecture is to add ``pseudo''-nodes to the network. 
Those nodes will not contribute to the network inference but will add noise to the network architecture. 
Figure~\ref{fig:hidden network example} shows an example of a simple network from Fiture~\ref{fig:example poly network} with two added pseudo-units: $PU11$ and $PU12$. 
To nullify the influence of the pseudo-units on the network inference the results of those units activation have to be canceled. 
A better way is to nullify an output edge weights, rather than input connection.
This way, we ensure that memory-enabled units or custom-activation units will not contribute. 
In the example, the edges: $PU11 \rightarrow U21$, $PU11 \rightarrow U22$, $PU12 \rightarrow U21$ and $PU12 \rightarrow U22$ will be zeroed.

The location of the units is randomized and the number of the units is depended on the need to hide the original network architecture.


\section{Communication-less MPC for Polynomial Calculations}
\label{sec:efficient MPC}
The goal of MPC calculations in the considered setup is to protect the published model from exposure to participating cloud providers. 
The model is trained by the data provider and has two components: architecture, which includes the layout, type, and interconnection of the neural units, as well as the weights of the input, which were refined during the training of the network, i.e. back-propagation phase.

Our goal is to protect the weights that were obtained by a costly process of training.
While the architecture also might hold ingenious insights, it is considered less of a secret and may be exposed to the cloud providers. 
In Section~\ref{sec:hiding architecture} we also discuss ways to hide the network architecture. 

Any MPC protocol can be used, preferably if it answers the following requirements. 
\begin{itemize}
    \item The protocol calculates polynomials over $k$ participating parties. Our goal is to spread the calculation over many servers/cloud providers to minimize the risk of adversaries' collaboration. Therefore, the protocol should preferably support $k > 2$ parties.
    \item Perfect Information theoretically secure protocol.
    \item Efficient calculation of the polynomial in terms of communication rounds to enable usage of high-degree polynomials. 
\end{itemize}

A number of MPC protocols answer those requirements~\cite{BenOr1988CompletenessTF,Chaum1988MultipartyUS}.
These MPC protocols based on Shamir secret sharing~\cite{10.1145/359168.359176} can cope with a minority of semi-honest parties and even with a third of the malicious parties. 
BGW protocol~\cite{BenOr1988CompletenessTF} provides a perfect security and~\cite{Chaum1988MultipartyUS} provides statistical security with any desirable certainty. 
In our case, the input is not a multi-variable that is secret-shared, but rather the weights and coefficients of the network are the secrets.\\

\noindent
{\bf Clear-text Inputs.}
In a simpler scenario, the input is revealed to all participating parties. 
In this case, the secrets are the weights of the trained network. 
The input values are then can be considered as numerical constants for the MPC calculation and thus, communication rounds can be eliminated completely, see BGW~\cite{BenOr1988CompletenessTF} algorithm where additive ``gates'' are calculated locally without any communication.  

Given a secret-share of coefficient $a$: $s=[s_1, s_2].$ 
The polynomial $p(x)$ can be calculated as $p(x) = p_1(x) + p_2(x),$ where $p_1(x)$ and $p_2(x)$ use the corresponding secret share.\\ 

\noindent
{\bf Secret-Shared Inputs.}
In the second scenario, the input values are protected as well, and thus, they are distributed by the secret share. 
As the input values are raised to polynomial degree $k$, the secret share is done on the set of values: $X=[x, x^2, \ldots x^k].$ 
Multiplication of secret shares requires communication rounds in a general case, still when secret sharing every element of $X$ it is possible to eliminate 
the communications all-together using ~\cite{DBLP:journals/iacr/BerendBD19}.

Notice that nested polynomials, Section~\ref{sec:nested polynomials}, cannot be used in this case, 
as the polynomial terms have to be regrouped for nesting, and secret-sharing of inputs prevents that.   

\section{Distributed Communication-less Secure Interference for Unknown DNN}
\label{sec:communicationless mpc for dnn}
The last two sections, Section~\ref{sec:multiple layers approximation} and Section~\ref{sec:efficient MPC}, 
provide all the required building blocks for communication-less MPC for common DNNs. 
In Section~\ref{sec:multiple layers approximation} we showed how a given, pre-trained network can be approximated with a single polynomial, in most common cases. 
As a side-note, as the neural network activation functions are not limited to a specific set, there might be networks that cannot be approximated.
However, the majority of networks use a rather small set of functions and architectures. 

Once the network is presented by a single polynomial, Section~\ref{sec:efficient MPC} shows that it can be calculated without a 
single communication round (apart from the input distribution and output gathering) when the inputs are revealed, 
or with half the communication rounds when the inputs are secret. 

Taken together, those two results enable a somewhat surprising outcome:
the data owner can train DNN models, pre-process, and share them with multiple cloud providers. 
The providers then can collaboratively calculate interference of the network on common or secret-shared inputs without ever communicating with each other. Thus, reducing the attack surface even further even for multi-layer networks.

\section{Fully Homomorphic Encryption Alternative}
\label{sec:FHA}
The reduction of Neural Networks to nested polynomials, facilitate inference over encrypted polynomial coefficients and encrypted inputs using computational secure (unlike the perfect information theoretic secure of the other scheme proposed here) Fully Homomorphic Encryption (FHE) \cite{Gentry:2009:FHE:1536414.1536440}. The nested polynomial that
represents fully connected layers can still be calculated in polynomial time
(the total number of connections is quadratic in the number of every two layers of neurons) so that some of the encrypted coefficients (or edge weights) can be an encrypted zero which in fact yields an (unrevealed) subset of the Neural Network. 

Delegation of machine learning to a third party (e.g., cloud provider) without revealing anything about the (big) data (collected), the inputs/queries, and the outputs is an important goal we address here, by using FHE and the nested polynomial. In certain cases, for example when the neuron computes the max function, the nested polynomial can integrate actual FHE computation of the max over the inputs arriving from the previous layer, rather than a polynomial over these inputs. 
A neuron is computed as polynomial over input polynomials (values), 
and two (or more) results can be computed for each neuron: one a polynomial over the inputs to the neuron and one an FHE max value over the input. 
Then use an encrypted bit(s) to blindly choose among the results, i.e. 
between polynomial or ``direct'' FHE calculation of the neuron activation function.

In the rest of the paper, we will show experiments performed to validate the results.


\section{Experiments}
\label{sec:experiments}
All tests were performed on the Fashion database of MNIST, which contains a training set of 60,000 and a testing set of 10,000 28x28 images of 10 fashion categories. 
The task is a multi-class classification of a given image. 

To solve the problem we have used a non-optimized neural network with two dense hidden layers: one of 300 units and the second one with 100 units. The output layer is a softmax layer with ten units and batch normalization layers before each activation layer.

The performed experiments were done on a pre-trained model. 
The model was loaded and translated into polynomial as described above automatically. 
This enables us to perform translation for any pre-trained network, similarly in spirit to \cite{kumar2019cryptflow}. Both the original model and polynomial representation were executed on the same inputs. The outputs are compared for different classification (divided by a total number of test inputs). 

\begin{figure}
    \label{fig:degree to accuracy}
    \centering
    \begin{minipage}{0.45\textwidth}
        \centering
        \includegraphics[width=0.9\textwidth]{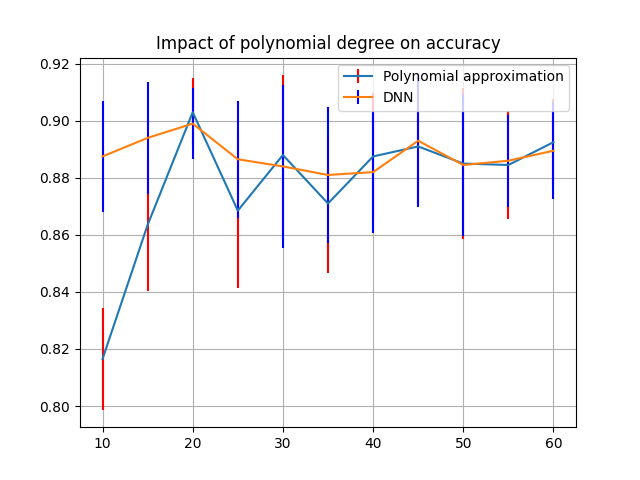} 
        \caption{Accuracy of DNN and polynomial approximation averaged over 10 runs of 500 examples each.}
    \end{minipage}\hfill
    \begin{minipage}{0.45\textwidth}
        \centering
        \includegraphics[width=0.9\textwidth]{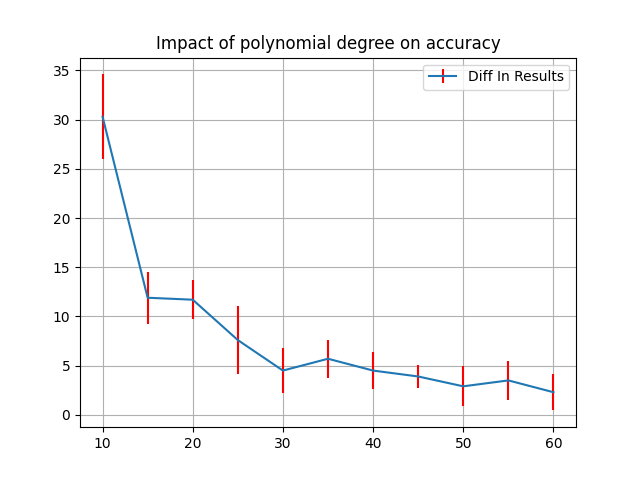} 
        \caption{Relative difference in results between polynomial approximation and the DNN model as a function of polynomial degree.}
    \end{minipage}
\end{figure}

Figure~\ref{fig:degree to accuracy} shows the difference in accuracy of the network with different degrees.
As can be seen, the accuracy improves with the degree of the polynomial approximation, however the improvement flattens at around $d=30$. 

The computation costs are increasing linearly with the polynomial degree (data not shown), where the original ReLu is similar to $d=1$ degree polynomial. Thus, it makes sense to choose the lowest degree that still provides consistent and accurate results. 

%

\section{Conclusions}
\label{sec:conclusions}
DNN networks are quickly becoming one of the most common computing workloads. 
Therefore, any optimization of their execution is important and could potentially save numerous CPU and network resources.
In this paper, we have presented a way to reduce and ultimately eliminate the number of communication rounds in the secure multi-party computation of DNN models. 
This setup of the computations is important for the secure sharing of DNN-based models. 
We think that this optimization method can enable more efficient DNN calculations and further progress in the process of privacy-preserving data sharing.
In particular, extra security and privacy factor is established, when no communication among the MPC participants is required, namely a participant cannot easily collude with other participants, as their identities are revealed to it. 

The above optimization of DNN evaluation targets the inference phase, which is done after the DNN-based model is shared and distributed across cloud providers. 
The network is not trained anymore, but only queried by the clients.
At this phase, the performance issues do not impact the data owners, which could be resource-limited end-devices, but rather are relevant for the cloud providers that have as much larger resources. 

DNNCoin (Deep Neural Network Coin), or SMCoin (Similitude Model Coin) can be structured in a distributed fashion similar Cryptocoins, where secret shares of the DNN coefficients are distributed among servers, possibly with a common Merkle tree root residing on Blockchain and proof for the share belonging to the Merkle tree. Such that a request for inference using a particular DNNCoin/SMCoin is accompanied with (cryptocurrency) payment to the owner of the Big Data used to create the DNNCoin/SMCoin, while limitig the number of queries to avoid revealing the (possibly, polynomial representing) DNNCoin/SMCoin.   

\bibliographystyle{abbrv}
\bibliography{biblio}

\end{document}